\documentclass[11pt]{article}
\usepackage{graphicx}
\usepackage{float}
\usepackage[a4paper, margin=2.5cm]{geometry}
\usepackage{amsmath,amssymb}
\usepackage{authblk}

\usepackage{titlesec}

\usepackage{physics}  

\titleformat{\section}{\normalfont\Large\bfseries}{\thesection.}{1em}{}

\title{Torsional Effects in the Coupling between Gravity and Spinors-
Yukawa Gravity}
\author{Elisa Varani\\Universit\`a Cattolica del Sacro Cuore, Milan (Italy)\\\texttt{elisa.varani@unicatt.it}}
\date{}
\usepackage{hyperref}
\begin{document}

\maketitle

\begin{abstract}
We study spinors in the framework of general relativity, starting from the Dirac field Lagrangian in the approximation of weak gravity. We focus on how fermions couple to gravity through the spin connection, and we analyze these couplings by analogy with the Ginzburg-Landau model and the Yukawa interaction known from the Higgs mechanism.

By solving the field equations, we explore how these couplings affect the spacetime metric. In particular, torsion generated by fermionic spin currents naturally emerges and leads to the breaking of Lorentz symmetry. As a consequence, gravity acquires a mass and fermions gain additional mass contributions through their interaction with this gravitational field. These effects are localized and diminish quickly with distance.

Our model offers an alternative explanation to phenomena usually attributed to dark matter and dark energy. We link these cosmological effects to chirality-flip processes of Majorana neutrinos interacting with a massive graviton. Right-handed Majorana neutrinos, which are sterile under Standard Model interactions, generate repulsive gravitational curvature and act as a source of dark energy, while left-handed neutrinos contribute to attractive gravitational effects akin to dark matter. The spin-gravity coupling modifies the curvature of spacetime, influencing galaxy rotation, the accelerated expansion of the universe, and the bending of light.

In short, the intrinsic spin of fermions, when coupled to gravity via torsion, changes gravity from a long-range, massless force to a short-range, massive one. This new framework provides fresh insights into fundamental physics and cosmology, potentially explaining dark matter and dark energy phenomena through spin-related gravitational effects.

 
\end{abstract}


\section{Introduction}
\subsection{Theoretical Framework and Relation to Einstein--Cartan Theory}

The model presented in this work extends the Einstein--Cartan--Sciama--Kibble framework by explicitly incorporating chiral spinor currents as active sources of spacetime torsion. While classical Einstein--Cartan theory already couples spinor fields to torsion via the antisymmetric part of the affine connection, it typically treats left- and right-handed components symmetrically.

In contrast, our formulation distinguishes between left-handed and right-handed spinor fields \( \psi_L, \psi_R \), introducing interaction terms of the form:
\begin{equation}
\varepsilon_{bck} \left( \bar{\psi}_L \sigma^k \psi_L - \bar{\psi}_R \sigma^k \psi_R \right)
\end{equation}
which induce local anisotropic and parity-violating perturbations in the metric tensor.

These chiral contributions break Lorentz and chiral symmetry at a local scale, generating mass-like effects for the gravitational field and for the fermions themselves. The model allows for spinor-induced gravitational screening, reminiscent of the Meissner effect in superconductivity, and predicts a flip process in which a left-handed neutrino transforms into a right-handed one through the emission of a graviton:
\begin{equation}
\nu_L \rightarrow h_{\mu\nu} + \nu_R
\end{equation}

This flipping mechanism, embedded in the torsional dynamics of spacetime, provides a geometric origin for fermionic mass and contributes to curvature modulation. Within this framework, left-handed neutrinos generate positive curvature (gravitational attraction), while right-handed neutrinos induce negative curvature (repulsion), suggesting a reinterpretation of dark matter and dark energy phenomena as effects of chiral gravitational interactions.

This chiral-torsional gravity model thus unifies gravitational mass generation, symmetry breaking, and cosmological dynamics through a spinor-based geometric mechanism beyond the Standard Model and General Relativity.

\subsection{Dirac Field in Weak Gravity}

The Lagrangian for the Dirac field in curved spacetime with weak gravitational effects is written as:
\begin{equation}
\mathcal{L}_f = \frac{i}{2} \bar{\psi} \gamma^\mu D_\mu \psi - \frac{i}{2} D_\mu \bar{\psi} \gamma^\mu \psi - m \bar{\psi} \psi
\end{equation}

The covariant derivatives of the spinor and its adjoint are given by:
\begin{equation}
D_\nu \psi = \partial_\nu \psi + \Gamma_\nu \psi
\end{equation}
\begin{equation}
D_\nu \bar{\psi} = \partial_\nu \bar{\psi} - \bar{\psi} \Gamma_\nu
\end{equation}

The spin connection \( \Gamma_\nu \) is defined as:
\begin{equation}
\Gamma_\nu = \frac{i}{4} e_b{}^\mu \nabla_\nu e_{a\mu} \, \hat{\sigma}^{ab}
\end{equation}

The covariant derivative of the tetrad is given by:
\begin{equation}
\nabla_\nu e_a{}^\mu = \partial_\nu e_a{}^\mu + \Gamma^\mu_{\nu\lambda} e_a{}^\lambda
\end{equation}

The Christoffel symbols \( \Gamma^\mu_{\nu\lambda} \) are defined as:
\begin{equation}
\Gamma^\mu_{\nu\lambda} = \frac{1}{2} g^{\mu\rho} \left( \partial_\nu g_{\rho\lambda} + \partial_\lambda g_{\rho\nu} - \partial_\rho g_{\nu\lambda} \right)
\end{equation}
We consider the weak gravity approximation, where the metric is expanded as:
\begin{equation}
g_{\mu\nu} = \eta_{\mu\nu} + h_{\mu\nu}
\end{equation}

The gamma matrices are corrected to first order by:
\begin{equation}
\gamma^\mu = \hat{\gamma}^\mu - \frac{1}{2} h^\mu_a\hat{\gamma}^a
\end{equation}

The spin connection simplifies to:
\begin{equation}
\Gamma^\mu \simeq \frac{i}{4} h^\mu _{c|b}\, \hat{\sigma}^{bc}
\end{equation}
where we define:
\begin{equation}
h^\mu_{c|b}{} = \partial_b h^\mu_c{}
\end{equation}

So, in linearized gravity the Lagrangian becomes:
\begin{align}
\mathcal{L}_f &= \frac{i}{2} \bar{\psi} \hat{\gamma}^\mu \partial_\mu \psi 
- \frac{i}{2} \partial_\mu \bar{\psi} \hat{\gamma}^\mu \psi 
- m \bar{\psi} \psi \nonumber \\
&\quad - \frac{1}{2} h_{\mu a} \hat{T}^{a\mu} 
- \frac{i}{4} h_{\mu a} h^\mu_{c|b}{} \cdot \frac{i}{4} \bar{\psi} \left\{ \hat{\gamma}^a, \hat{\sigma}^{bc} \right\} \psi \nonumber \\
&\quad - \frac{i}{4} h_{\mu c|b} \cdot \frac{i}{4} \bar{\psi} \left\{ \hat{\gamma}^\mu, \hat{\sigma}^{bc} \right\} \psi
\end{align}

We denote the Lagrangian as the sum of a flat-spacetime part and a gravitational correction:
\begin{equation}
\mathcal{L}_f = \mathcal{L}_{\text{FLAT}} + \mathcal{L}_{Gf}
\end{equation}

Here, \( \mathcal{L}_{Gf} \) represents the adjunctive terms arising in weak gravity. It includes interaction terms between fermions and the weak gravitational field \( h_{\mu\nu} \), and also contributions from the spin connection structure.

We can draw a similarity with the Yukawa interaction:
\begin{equation}
\mathcal{L}_Y = -k \bar{\psi} \phi \psi
\end{equation}
where \( k \) is the Yukawa coupling constant and \( \phi \) is a scalar bosonic field.

In this analogy, \( h_{\mu\nu} \) plays a role similar to that of \( \phi \), mediating mass-like interactions with the spinor field through geometric (torsional) couplings.
\subsection{Spin connection terms}

Without the spin connection, the coupling between the gravitational field and fermions is written as
\begin{equation}
    \widehat{\mathcal{L}}_{Gf} = -\frac{1}{2} h^{\mu}_a \, \widehat{T}^{a}_{\mu}
\end{equation}
where the spinor stress-energy tensor in flat spacetime is given by
\begin{equation}
    \widehat{T}^{a}_{\mu} = \frac{i}{2} \bar{\psi} \widehat{\gamma}^{a} \partial_{\mu} \psi - \frac{i}{2} \partial_{\mu} \bar{\psi} \widehat{\gamma}^{a} \psi
\end{equation}

Including the contributions from the spin connection, the gravitational-fermion Lagrangian becomes
\begin{equation}
    \mathcal{L}_{Gf} = -\frac{1}{2} h^{\mu}_a \, \widehat{T}^{a}_{\mu} 
    + \frac{i}{2} h_{\mu c|b} \cdot \frac{i}{4} \left[\bar{\psi} \left\{ \widehat{\gamma}^{\mu}, \widehat{\sigma}^{bc} \right\} \psi \right] 
    - \frac{i}{4} h^{\mu}_a h_{\mu c|b} \cdot \frac{i}{4} \left[\bar{\psi} \left\{ \widehat{\gamma}^{a}, \widehat{\sigma}^{bc} \right\} \psi \right]
\end{equation}

We identify the spin connection contribution as:
\begin{equation}
    s^{abc} = -\frac{i}{4} \left[\bar{\psi} \left\{ \widehat{\gamma}^{a}, \widehat{\sigma}^{bc} \right\} \psi \right]
\end{equation}
so that the Lagrangian takes the form:
\begin{equation}
    \mathcal{L}_{Gf} = -\frac{1}{2} h^{\mu}_a \, \widehat{T}^{a}_{\mu} - \frac{i}{2} h_{\mu c|b} \, s^{\mu bc} + \frac{i}{4} h^{\mu}_a h_{\mu c|b} \, s^{abc}
\end{equation}

In explicit spinorial terms:
\begin{equation}
    \mathcal{L}_{Gf} = -\frac{1}{2} h^{\mu}_a \, \widehat{T}^{a}_{\mu} 
    - \frac{1}{8}h_{\mu c|b}\left[\bar{\psi} \left\{ \widehat{\gamma}^{\mu}, \widehat{\sigma}^{bc} \right\} \psi \right]
    + \frac{1}{16} h^{\mu}_a h_{\mu c|b} \left[\bar{\psi} \left\{ \widehat{\gamma}^{a}, \widehat{\sigma}^{bc} \right\} \psi \right]
\end{equation}

We compare these terms to the standard Yukawa interaction:
\begin{equation}
    \mathcal{L}_Y = -k \bar{\psi} \, \phi \, \psi
\end{equation}

In our model, the role of the boson field \( \phi \) is effectively played by a gravitational structure filtered through the matrix combination \( \left\{ \widehat{\gamma}^{a}, \widehat{\sigma}^{bc} \right\} \). We note that the spinor angular momentum contributes to a Yukawa-like interaction mediated by this gamma-matrix structure. 

The gravitational field enters this interaction only if the derivative \( h_{\mu c|b}\neq 0 \), implying that the interaction is stronger when the gravitational field is dynamical.

Torsion is therefore the geometrical mediator of the spinor-gravity interaction. At first order, we investigate:
\begin{equation}
    \mathcal{L}_I' = -\frac{i}{2} h_{\mu c|b} s^{\mu bc}
\end{equation}
At second order:
\begin{equation}
    \mathcal{L}_I'' = \frac{i}{4} h^{\mu}_a h_{\mu c|b} s^{abc}
\end{equation}

Thus, the full interaction terms can be expressed as:
\begin{equation}
    \mathcal{L}_I = -\frac{1}{8} h_{\mu c|b} \left[\bar{\psi} \left\{ \widehat{\gamma}^{\mu}, \widehat{\sigma}^{bc} \right\} \psi \right]
    + \frac{1}{16} h^{\mu}_a h_{\mu c|b}\left[\bar{\psi} \left\{ \widehat{\gamma}^{a}, \widehat{\sigma}^{bc} \right\} \psi \right]
\end{equation}

In the standard Yukawa model, the mass of fermions arises from the interaction with a scalar boson:
\begin{equation}
    \mathcal{L}_Y = -k \bar{\psi} \, \phi \, \psi
\end{equation}

In this spinor-gravity model, the mass-like interaction is realized via torsion. One can argue, in analogy with the Higgs mechanism, that the coupling between spinors and torsion leads to a form of “mass generation,” observable through gravitational lensing.

This reinforces our previous results, indicating that torsion contributes to the cosmological constant and can drive the expansion of the universe, potentially replacing or supplementing the role of dark matter [1].

\section{Material and Methods}
\subsection{Ginzburg–Landau Mechanism and Higgs Model}

A comparison is drawn between this model and two foundational frameworks of spontaneous symmetry breaking: the Ginzburg–Landau theory, characterized by \( U(1) \) symmetry breaking, and the Higgs mechanism, involving the breaking of the \( SU(2) \times U(1) \) gauge symmetry.

In both models, the potential is of the form:
\begin{equation}
V(\phi) = \mu^2 \phi^2 + \lambda \phi^4,
\end{equation}
with \( \mu^2 < 0 \) and \( \lambda > 0 \).  
Here, \( \phi \) represents a bosonic field: in the Ginzburg–Landau theory it corresponds to the probability amplitude of condensed Cooper pairs, while in the Higgs model \( \phi \) denotes the Higgs boson field.

The Lagrangian for the Ginzburg–Landau model is:
\begin{equation}
\mathcal{L}_\phi = \frac{1}{4} \partial_\mu \phi \, \partial^\mu \phi - \mu^2 \phi^2 - \lambda \phi^4.
\end{equation}

The total Lagrangian then takes the form:
\begin{equation}
\mathcal{L}_{\text{tot}} = \mathcal{L}_f + \mathcal{L}_\phi + \mathcal{L}_y,
\end{equation}
where the Yukawa interaction is:
\begin{equation}
\mathcal{L}_y = -k \, \bar{\psi} \, \phi \, \psi,
\end{equation}
and the fermion term in flat spacetime is:
\begin{equation}
\mathcal{L}_f = \frac{i}{2} \bar{\psi} \gamma^\mu \partial_\mu \psi - \frac{i}{2} \partial_\mu \bar{\psi} \gamma^\mu \psi - m \bar{\psi} \psi.
\end{equation}

Now, we build the analog interaction using gravity and torsion. In the weak gravity approximation, we obtain a Yukawa-like interaction:
\begin{equation}
\mathcal{L}_I = -\frac{1}{8} h_{\mu c|b}\, \left[ \bar{\psi} \{ \hat{\gamma}^\mu, \hat{\sigma}^{bc} \} \psi \right]
+ \frac{1}{16} h_a^\mu h_{\mu c|b} \, \left[ \bar{\psi} \{ \hat{\gamma}^a, \hat{\sigma}^{bc} \} \psi \right],
\end{equation}
where the spin connection terms mediate a torsion-based coupling.

We first analyze the term:
\begin{equation}
V_1(h_{\mu c|b}\{ \hat{\gamma}^\mu, \hat{\sigma}^{bc} \}) = \mu^2 \left(h_{\mu c|b}\{ \hat{\gamma}^\mu, \hat{\sigma}^{bc} \}\right)^2 + \lambda \left(h_{\mu c|b}\{ \hat{\gamma}^\mu, \hat{\sigma}^{bc} \}\right)^4.
\end{equation}
The corresponding kinetic and potential Lagrangian is:
\begin{equation}
\mathcal{L}_h = \frac{1}{4} \partial_\nu h_{\mu c|b} \, \partial^\nu h_{\mu c|b}\, \left( \{ \hat{\gamma}^\nu, \hat{\sigma}^{bc} \} \right)^2 - V_1.
\end{equation}

We then analyze the second-order term:
\begin{equation}
V_2(h_a^\mu h_{\mu c|b}\{ \hat{\gamma}^a, \hat{\sigma}^{bc} \}) = \mu^2 \left(h_a^\mu h_{\mu c|b} \{ \hat{\gamma}^a, \hat{\sigma}^{bc} \}\right)^2 + \lambda \left(h_a^\mu h_{\mu c|b}\{ \hat{\gamma}^a, \hat{\sigma}^{bc} \}\right)^4,
\end{equation}
with the associated Lagrangian:
\begin{equation}
\mathcal{L}_{hh} = \frac{1}{4} \partial_\nu (h_a^\mu h_{\mu c|b}) \, \partial^\nu (h_a^\mu h_{\mu c|b}) \left( \{ \hat{\gamma}^a, \hat{\sigma}^{bc} \} \right)^2 - V_2.
\end{equation}

In the weak gravity limit, the total Lagrangian becomes:
\begin{equation}
\mathcal{L}_{\text{tot}} = \mathcal{L}_f + \mathcal{L}_h + \mathcal{L}_{hh},
\end{equation}
where \( \mathcal{L}_f = \mathcal{L}_{\text{FLAT}} + \mathcal{L}_{Gf} \).

These terms \( \mathcal{L}_h + \mathcal{L}_{hh} \) represent the interaction between torsion and gravity. Torsion is induced by spinor fields, and it enables the coupling between fermions and the gravitational field. This mechanism parallels the Higgs model, where fermions couple via bosonic mediators.

Thus, in this formulation:
\begin{equation}
h_{\mu c|b}\left[ \bar{\psi} \{ \hat{\gamma}^\mu, \hat{\sigma}^{bc} \} \psi \right]
\end{equation}
plays the role of the Yukawa interaction, and
\begin{equation}
h_{\mu c|b}\{ \hat{\gamma}^\mu, \hat{\sigma}^{bc} \}
\end{equation}
effectively replaces the Higgs field \( \phi \) as the interaction mediator.

\subsection{Gravitational Field and Symmetry Breaking: Torsion Coupling}

We investigate the fields \( \tilde{\phi} = h_{\mu c|b}\{ \hat{\gamma}^\mu, \hat{\sigma}^{bc} \} \) and \( \tilde{\phi} = h_a^\mu h_{\mu c|b} \{ \hat{\gamma}^a, \hat{\sigma}^{bc} \} \), analogues of the Higgs boson \( \phi \) or Cooper condensate in superconductor theory. The combination \( \{ \hat{\gamma}^\mu, \hat{\sigma}^{bc} \} \) defines the interaction structure between tetrad perturbations and spinor currents, effectively forming the "grid" through which torsion interacts with fermions.

The corresponding kinetic and potential terms in the Lagrangian take the form:

\begin{equation}
\mathcal{L}_h = \frac{1}{4} \partial_\nu \left(h_{\mu c|b} \right) \partial^\nu \left(h_{\mu c|b} \right) \left( \{ \hat{\gamma}^\mu, \hat{\sigma}^{bc} \} \right)^2 - \mu^2 \left( h_{\mu c|b} \{ \hat{\gamma}^\mu, \hat{\sigma}^{bc} \} \right)^2 - \lambda \left( h_{\mu c|b}\{ \hat{\gamma}^\mu, \hat{\sigma}^{bc} \} \right)^4.
\end{equation}

Assuming \( \mu^2 < 0 \) and \( \lambda > 0 \), the potential develops a non-zero vacuum expectation value. We calculate the minimum of the potential:

\begin{equation}
V\left( h_{\mu c|b} \{ \hat{\gamma}^\mu, \hat{\sigma}^{bc} \} \right) = \mu^2 \left( h_{\mu c|b} \{ \hat{\gamma}^\mu, \hat{\sigma}^{bc} \} \right)^2 + \lambda \left(h_{\mu c|b} \{ \hat{\gamma}^\mu, \hat{\sigma}^{bc} \} \right)^4.
\end{equation}

We obtain a minimum different from zero:

\begin{equation}
 h_{\mu c|b}\{ \hat{\gamma}^\mu, \hat{\sigma}^{bc} \} = \pm \sqrt{\frac{-\mu^2}{2\lambda}}.
\end{equation}

This spontaneous symmetry breaking is associated with the presence of a fermionic current, which naturally arises in general relativity when spinors are included.

We calculate the minimum for the second potential with \( \alpha^2 < 0 \) and \( \beta > 0 \):

\begin{equation}
V\left( h_a^\mu h_{\mu c|b} \{ \hat{\gamma}^a, \hat{\sigma}^{bc} \} \right) = \alpha^2 \left( h_a^\mu h_{\mu c|b} \{ \hat{\gamma}^a, \hat{\sigma}^{bc} \} \right)^2 + \beta \left( h_a^\mu h_{\mu c|b} \{ \hat{\gamma}^a, \hat{\sigma}^{bc} \} \right)^4.
\end{equation}

We obtain:

\begin{equation}
h_a^\mu h_{\mu c|b}\{ \hat{\gamma}^\mu, \hat{\sigma}^{bc} \} = \pm \sqrt{\frac{-\alpha^2}{2\beta}}.
\end{equation}

In summary, the gravitational field emerges from the equations of symmetry breaking and torsion-induced modification of gravitational coupling. The interaction between spinors and the gravitational field, mediated by torsion, alters the structure of spacetime, introducing a curvature that can no longer be described through the coupling between the metric and matter alone, but through an extended coupling that also includes torsion. The latter is associated with the non-zero minimum of a potential, which — similarly to the Ginzburg-Landau mechanism — arises from the interaction between spinors and the gravitational field.

The breaking of Lorentz symmetry that occurs in this context involves the acquisition of a gravitational mass. Gravity thus behaves as a massive and short-range interaction, undergoing a shielding effect in the affected areas.

The metric perturbation \( h_{\mu\nu} \), which emerges in the model, is directly associated with the torsion and modification of the gravitational field due to the interaction with spinors. In particular, in vortex regions where spinorial currents are concentrated, gravity is expelled, generating a gravitational shielding phenomenon. This mechanism is analogous to the Meissner effect in superconductors, where the magnetic field is ejected from the material. In cosmology, such an effect could explain the accelerated expansion of the universe without resorting to dark matter or dark energy.

Finally, the breaking of Lorentz symmetry implies a modification of geodesic trajectories. The deviation between nearby geodesics, typically described in general relativity by the curvature tensor, is altered by the presence of torsion. In particular, the repulsive gravitational effect — the ejection of gravity from swirling regions dominated by fermionic spin currents — could give rise to spiral-shaped trajectories, similar to those observed in galactic dynamics and certain gravitational lensing phenomena. A more detailed discussion of geodesic deviation in the presence of torsion and its dynamic implications will be developed in a subsequent work.
\subsection{Full Lagrangian and Field Equations}

We consider the total Lagrangian \(L_{\rm tot}\) which describes the interaction between fermions and gravity, including torsional effects due to fermionic currents.  It is the sum of three components:
\begin{equation}
L_{\rm tot} \;=\; L_f \;+\; L_h \;+\; L_{hh}\,,
\end{equation}
where \(L_f\) is the fermion Lagrangian, \(L_h\) the gravitational–torsion perturbation Lagrangian, and \(L_{hh}\) the second‐order metric perturbation Lagrangian.  We write
\begin{equation}
L_f \;=\; L_{\rm flat} + L_{Gf}\,,
\end{equation}
with
\begin{align}
L_f &= \frac{i}{2}\,\bar\psi\,\hat\gamma^\mu\partial_\mu\psi 
         - \frac{i}{2}\,\partial_\mu\bar\psi\,\hat\gamma^\mu\psi 
         - m\,\bar\psi\psi
         - \frac{1}{2}\,h_{\mu a}\,\hat T^{a\mu}\nonumber\\
     &\quad - \frac{i}{4}\,h_{\mu a}\,h_{\mu c|b}\,
         \bigl[\bar\psi\{\hat\gamma^a,\hat\sigma^{bc}\}\psi\bigr]
       - \frac{i}{4}\,h_{\mu\,c|b}\,
         \bigl[\bar\psi\{\hat\gamma^\mu,\hat\sigma^{bc}\}\psi\bigr],
\end{align}
where \(\hat T^{a\mu}=\tfrac{i}{2}\bar\psi\hat\gamma^a\partial^\mu\psi-\tfrac{i}{2}\partial^\mu\bar\psi\hat\gamma^a\psi\).

The gravitational perturbation Lagrangian is
\begin{equation}
L_h
= \frac{1}{4}\,\partial_\nu\!h_{\mu\,c|b}\,\partial^\nu\!h_{\mu\,c|b}\,\{\hat\gamma^\mu,\hat\sigma^{bc}\}^2
- \mu^2 \bigl(h_{\mu\,c|b}\{\hat\gamma^\mu,\hat\sigma^{bc}\}\bigr)^2
- \lambda \bigl(h_{\mu\,c|b}\{\hat\gamma^\mu,\hat\sigma^{bc}\}\bigr)^4\,,
\end{equation}
and the second‐order term is
\begin{equation}
L_{hh}
= \frac{1}{4}\,\partial_\nu\bigl(h_a^\mu h_{\mu\,c|b}\bigr)\partial^\nu\bigl(h_a^\mu h_{\mu\,c|b}\bigr)\{\hat\gamma^a,\hat\sigma^{bc}\}^2
- \alpha^2\bigl(h_a^\mu h_{\mu\,c|b}\{\hat\gamma^a,\hat\sigma^{bc}\}\bigr)^2
- \beta \bigl(h_a^\mu h_{\mu\,c|b}\{\hat\gamma^a,\hat\sigma^{bc}\}\bigr)^4.
\end{equation}

Combining all terms, the full Lagrangian reads
\begin{align}
L_{\rm tot}
&= \frac{i}{2}\,\bar\psi\,\hat\gamma^\mu\partial_\mu\psi 
  - \frac{i}{2}\,\partial_\mu\bar\psi\,\hat\gamma^\mu\psi 
  - m\,\bar\psi\psi
  - \frac{1}{2}\,h_{\mu a}\,\hat T^{a\mu} \nonumber\\
&\quad + \frac{1}{8}\,h_{\mu a}\,h_{\mu c|b}\!
         \bigl[\bar\psi\{\hat\gamma^a,\hat\sigma^{bc}\}\psi\bigr]
       + \frac{1}{8}\,h_{\mu\,c|b}\!
         \bigl[\bar\psi\{\hat\gamma^\mu,\hat\sigma^{bc}\}\psi\bigr] \nonumber\\
&\quad + \frac{1}{4}\,\partial_\nu\!h_{\mu\,c|b}\,\partial^\nu\!h_{\mu\,c|b}\,\{\hat\gamma^\mu,\hat\sigma^{bc}\}^2
  - \mu^2 \bigl(h_{\mu\,c|b}\{\hat\gamma^\mu,\hat\sigma^{bc}\}\bigr)^2
  - \lambda \bigl(h_{\mu\,c|b}\{\hat\gamma^\mu,\hat\sigma^{bc}\}\bigr)^4 \nonumber\\
&\quad + \frac{1}{4}\,\partial_\nu\bigl(h_a^\mu h_{\mu\,c|b}\bigr)\partial^\nu\bigl(h_a^\mu h_{\mu\,c|b}\bigr)\{\hat\gamma^a,\hat\sigma^{bc}\}^2
  - \alpha^2\bigl(h_a^\mu h_{\mu\,c|b}\{\hat\gamma^a,\hat\sigma^{bc}\}\bigr)^2
  - \beta \bigl(h_a^\mu h_{\mu\,c|b}\{\hat\gamma^a,\hat\sigma^{bc}\}\bigr)^4.
\end{align}
\section{Results}
\subsection{ Field Solutions}

The Lagrangian for $h_{\mu c|b}$ is given by:
\begin{align}
\mathcal{L}_{h_{\mu c|b}} = &+\frac{1}{8} h_{\mu a} h_{\mu c|b}\left[\bar{\psi} \{ \hat{\gamma}^a, \hat{\sigma}^{bc} \} \psi \right] + \frac{1}{8} h_{\mu c|b} \left[\bar{\psi} \{ \hat{\gamma}^\mu, \hat{\sigma}^{bc} \} \psi \right] \nonumber \\
& + \frac{1}{4} \partial_\nu h_{\mu c|b} \, \partial^\nu h_{\mu c|b} \, \{ \hat{\gamma}^\mu, \hat{\sigma}^{bc} \}^2 
- \mu^2 \left(h_{\mu c|b} \{ \hat{\gamma}^\mu, \hat{\sigma}^{bc} \} \right)^2 
- \lambda \left(h_{\mu c|b} \{ \hat{\gamma}^\mu, \hat{\sigma}^{bc} \} \right)^4 \nonumber \\
& + \frac{1}{4} \partial_\nu (h_a^\mu h_{\mu c|b}) \, \partial^\nu (h_a^\mu h_{\mu c|b}) \{ \hat{\gamma}^a, \hat{\sigma}^{bc} \}^2 
- \alpha^2 \left(h_a^\mu h_{\mu c|b} \{ \hat{\gamma}^a, \hat{\sigma}^{bc} \} \right)^2 
- \beta \left(h_a^\mu h_{\mu c|b} \{ \hat{\gamma}^a, \hat{\sigma}^{bc} \} \right)^4.
\end{align}

We simplify the Lagrangian by omitting the term $\frac{1}{8} h_{\mu a} h_{\mu c|b} \left[\bar{\psi} \{ \hat{\gamma}^a, \hat{\sigma}^{bc} \} \psi \right]$ and higher-order terms in $\lambda$:
\begin{equation}
\mathcal{L}_{h_{\mu c|b}} \approx \frac{1}{4} \partial_\nu h_{\mu c|b} \, \partial^\nu h_{\mu c|b} \, \{ \hat{\gamma}^\mu, \hat{\sigma}^{bc} \}^2 - \mu^2 \left(h_{\mu c|b} \{ \hat{\gamma}^\mu, \hat{\sigma}^{bc} \} \right)^2 + \frac{1}{8} h_{\mu c|b} \left[\bar{\psi} \{ \hat{\gamma}^\mu, \hat{\sigma}^{bc} \} \psi \right].
\end{equation}

The field equations are obtained as:
\begin{align}
\frac{\partial \mathcal{L}}{\partial h_{\mu c|b}} &= \frac{1}{8} \left[\bar{\psi} \{ \hat{\gamma}^\mu, \hat{\sigma}^{bc} \} \psi \right] - 2\mu^2 h_{\mu c|b} \left(\{ \hat{\gamma}^\mu, \hat{\sigma}^{bc} \} \right)^2, \\
\frac{\partial \mathcal{L}}{\partial (\partial_\nu h_{\mu c|b})} &= \frac{1}{4} \partial^\nu h_{\mu c|b} \, \{ \hat{\gamma}^\mu, \hat{\sigma}^{bc} \}^2, \\
\Rightarrow \quad \partial_\nu \partial^\nu h_{\mu c|b} \, \{ \hat{\gamma}^\mu, \hat{\sigma}^{bc} \}^2 &= 2 \bar{\psi} \{ \hat{\gamma}^\mu, \hat{\sigma}^{bc} \} \psi - 8 \mu^2 h_{\mu c|b} \left( \{ \hat{\gamma}^\mu, \hat{\sigma}^{bc} \} \right)^2.
\end{align}

This leads to the wave equation with two sources: the fermionic current and the mass term. Assuming $\mu^2 < 0$, we define $\mu^2 = -m^2$:
\begin{equation}
\partial_\nu \partial^\nu h_{\mu c|b} - 8m^2 h_{\mu c|b} = S(x),
\end{equation}
with source term:
\begin{equation}
S(x) = \frac{2 \bar{\psi} \{ \hat{\gamma}^\mu, \hat{\sigma}^{bc} \} \psi}{\left( \{ \hat{\gamma}^\mu, \hat{\sigma}^{bc} \} \right)^2}.
\end{equation}

The homogeneous equation reads:
\begin{equation}
\partial_\nu \partial^\nu h_{\mu c|b} - 8m^2 h_{\mu c|b} = 0,
\end{equation}
with plane-wave solution:
\begin{equation}
h_{\mu c|b}(x) = A e^{i k_\nu x^\nu} + B e^{-i k_\nu x^\nu}, \quad \text{with} \quad k_\nu k^\nu = 8m^2.
\end{equation}

This is a Klein-Gordon-like equation for a massive field.

The solution with source is:
\begin{equation}
h_{\mu c|b}(x) = \int d^4x' \, G(x,x') \, \frac{2 \bar{\psi}(x') \{ \hat{\gamma}^\mu, \hat{\sigma}^{bc} \} \psi(x')}{\left( \{ \hat{\gamma}^\mu, \hat{\sigma}^{bc} \} \right)^2},
\end{equation}
where the propagator is:
\begin{equation}
G(x,x') = \int \frac{d^4k}{(2\pi)^4} \, \frac{e^{-i k_\nu (x^\nu - x'^\nu)}}{k_\nu k^\nu - 8m^2 + i\epsilon}.
\end{equation}
\begin{figure}[H]
    \centering
    \includegraphics[width=0.7\textwidth]{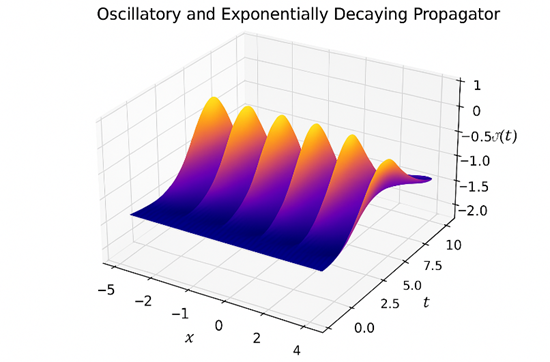}
    \caption{Green function.}
    \label{fig:etichetta}
\end{figure}
The full solution is:
\begin{equation}
\partial_b h_{\mu c}(x) = h_{\mu c|b}(x) = A_{\mu c} e^{i k_\nu x^\nu} + B_{\mu c} e^{-i k_\nu x^\nu} + \int d^4x' \, G(x,x') \, \frac{2 \bar{\psi}(x') \{ \hat{\gamma}^\mu, \hat{\sigma}^{bc} \} \psi(x')}{\left( \{ \hat{\gamma}^\mu, \hat{\sigma}^{bc} \} \right)^2}.
\end{equation}

For a static approximation:
\begin{equation}
G(x,x') \approx \frac{e^{-8 m^2 |x - x'|}}{|x - x'|},
\end{equation}
indicating exponential decay due to the mass term.

Assuming local Lorentz symmetry breaking in a Planck-scale region:
\[
|x - x'| = l_p = \sqrt{\frac{\hbar G}{c^3}}, \quad m \approx m_p = \frac{\hbar}{l_p c},
\]
we get:
\begin{equation}
G(x,x') \approx \frac{e^{-8 m_p^2 l_p}}{l_p} = \frac{e^{-8}}{l_p},
\end{equation}
indicating strong suppression of the propagator at Planck scale.

The final expression for $h_{\mu c}(x)$ is:
\begin{equation}
h_{\mu c}(x) = \int dx_b \left( A_{\mu c} e^{i k_\nu x^\nu} + B_{\mu c} e^{-i k_\nu x^\nu} + \int d^4x' \, G(x,x') \, \frac{2 \bar{\psi}(x') \{ \hat{\gamma}^\mu, \hat{\sigma}^{bc} \} \psi(x')}{\left( \{ \hat{\gamma}^\mu, \hat{\sigma}^{bc} \} \right)^2} \right).
\end{equation}

\noindent In summary, the equation shows that the gravitational field perturbation \(h_{\mu c}(x)\) has two components:
\begin{itemize}
  \item A wave-like solution that propagates as a massive field due to torsional effects causing Lorentz symmetry breaking.
  \item A particular solution that is sourced by the fermionic spin current. This interaction is short-ranged due to the massive nature of the gravitational field.
\end{itemize}
\noindent The field equations suggest that the interaction between gravity and spinors through torsion leads to modifications of gravity—such as the emergence of an effective mass and screening effects—which could potentially explain phenomena usually attributed to dark matter. The source term in the particular solution explicitly links the gravitational field to the presence and properties of fermions, highlighting the central role of the spinor field and torsion in this model.[2]

\noindent We approximate a solution with a central singularity oscillating, from positive to negative value.
\begin{figure}[H]
    \centering
    \includegraphics[width=0.5\textwidth]{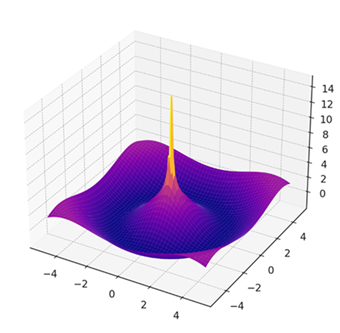}
    \caption{Central singularity}
    \label{fig:etichetta}
\end{figure}

\subsection{Specific Fields Solution with Weyl Spinors}

For further analysis we make explicit the torsional elements in the spin current considering the matrix
\begin{equation}
M^{\mu bc} = \{\hat\gamma^{\mu}, \hat\sigma^{bc}\}, 
\qquad
\hat\gamma^\mu = e_a^\mu \hat\gamma^a = \bigl(\delta_a^\mu - \tfrac{1}{2}h_a^\mu\bigr)\,\hat\gamma^a.
\end{equation}
In flat space and to first order,
\begin{equation}
\hat\gamma^\mu \approx \hat\gamma^a.
\end{equation}
The only non-vanishing components of \(M^{abc} = \{\hat\gamma^a,\hat\sigma^{bc}\}\) are
\begin{align}
M^{0bc} &= 2\,\varepsilon_{bck}
  \begin{pmatrix}\sigma^k & 0 \\ 0 & -\sigma^k\end{pmatrix}, \label{eq:M0bc}\\
M^{a0c} &= -M^{0ac} = -2\,\varepsilon_{ack}
  \begin{pmatrix}\sigma^k & 0 \\ 0 & -\sigma^k\end{pmatrix}, \label{eq:Ma0c}\\
M^{kbc} &= 2\,\varepsilon_{bck}
  \begin{pmatrix}0 & I \\ -I & 0\end{pmatrix}. \label{eq:Mkbc}
\end{align}

We now consider only the particular solution with the spin-current source:
\begin{equation}
h_{ac}(x)
= \int \mathrm{d}x_b 
  \biggl(\int \mathrm{d}^4x'\,G(x,x')\,
    \frac{2\,\bar\psi\{\hat\gamma^a,\hat\sigma^{bc}\}\psi}
         {\{\hat\gamma^a,\hat\sigma^{bc}\}^2}\biggr).
\end{equation}
For Weyl spinors \(\psi = \begin{pmatrix}\psi_L \\ \psi_R\end{pmatrix}\), one finds
\begin{equation}
2\,\bar\psi\,M^{0bc}\,\psi
= 4\,\varepsilon_{bck}\bigl(\bar\psi_L\sigma^k\psi_L
     - \bar\psi_R\sigma^k\psi_R\bigr),
\quad
(M^{0bc})^2 = 12\,I.
\end{equation}
Hence, for the time-like component,
\begin{equation}
h_{0c}(x)
= \frac{1}{3}\int \mathrm{d}x_b 
  \biggl(\int \mathrm{d}^4x'\,G(x,x')\,
    \varepsilon_{bck}\bigl(\bar\psi_L\sigma^k\psi_L
     - \bar\psi_R\sigma^k\psi_R\bigr)\biggr).
\end{equation}
And for the space-like case,
\begin{equation}
h_{ac}(x)
= -\frac{1}{3}\int \mathrm{d}x^0 
  \biggl(\int \mathrm{d}^4x'\,G(x,x')\,
    \varepsilon_{ack}\bigl(\bar\psi_L\sigma^k\psi_L
     - \bar\psi_R\sigma^k\psi_R\bigr)\biggr).
\end{equation}

For space-like indices \(a,b,c\),
\begin{equation}
2\,\bar\psi\,M^{abc}\,\psi
= 4\,\varepsilon_{bca}\bigl(\bar\psi_L\psi_R
     - \bar\psi_R\psi_L\bigr),
\quad
(M^{abc})^2 = -12\,I,
\end{equation}
giving
\begin{equation}
h_{ac}(x)
= -\frac{1}{3}\int \mathrm{d}x_b
  \biggl(\int \mathrm{d}^4x'\,G(x,x')\,
    \varepsilon_{bca}\bigl(\bar\psi_L\psi_R
     - \bar\psi_R\psi_L\bigr)\biggr).
\end{equation}

The approximated solution for \(h_{\mu c}(x)\) exhibits short-distance decay from the oscillating peak. Negative values correspond to repulsive (anti-gravitational) regions. Fermion spin thus acts like a harmonic oscillator for gravity—alternating attraction and repulsion—shaping torsion in spacetime.
\begin{figure}[H]
    \centering
    \includegraphics[width=0.5\textwidth]{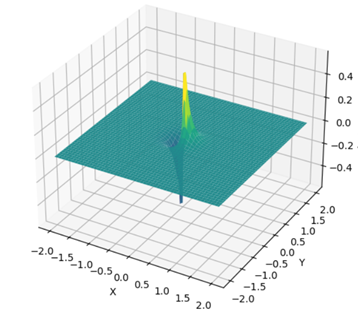}
    \caption{Oscillating peak.}
    \label{fig:etichetta}
\end{figure}

If the metric oscillates between positive and negative values, regions of converging (gravity) and diverging (antigravity) curvature emerge. Chiral terms drive this effect, linking chiral symmetry breaking to antigravity. In a universe dominated by left-handed fermions, gravity appears attractive; fluctuations favoring right-handed components induce negative curvature and repulsion typical of antigravity.[3]
\subsection{Chirality affects Gravity}

Consider the relationship between chirality and perturbations of the gravitational field. The two terms are not symmetric with respect to chiral symmetry: 

\begin{equation}
(\bar{\psi}_L \sigma^k \psi_L - \bar{\psi}_R \sigma^k \psi_R)
\end{equation}

\begin{equation}
(\bar{\psi}_L \psi_R - \bar{\psi}_R \psi_L)
\end{equation}

The first current maintains chiral symmetry while violating parity. The left-handed component is independent of the right-handed one. The second term, on the other hand, also violates chiral symmetry and is antisymmetric with respect to parity. The Yukawa interaction is a classic example of an interaction that breaks chiral symmetry because it mixes left-handed and right-handed fermions via a Higgs field and generates a mass term that couples the two components.

The Yukawa interaction is given by:

\begin{equation}
L_y = -k(\bar{\psi}_L \varphi \psi_R - \bar{\psi}_R \varphi \psi_L)
\end{equation}

The mass term that emerges from the Yukawa interaction, after the Higgs field has acquired a non-zero value, is:

\begin{equation}
m(\bar{\psi}_L \psi_R + \bar{\psi}_R \psi_L)
\end{equation}

In weak gravity, the curvature tensor takes the simplified form and only holds the term \( \Box h_{\mu \nu}(x) \) to describe the dynamic propagation of gravitational perturbations in a weak gravity context:

\begin{equation}
R_{\mu \nu} = -\frac{1}{2} \eta^{\rho \sigma} \partial_\rho \partial_\sigma h_{\mu \nu}(x) = -\frac{1}{2} \Box h_{\mu \nu}(x)
\end{equation}

Let's clarify the use of flat indices and world indices:

\begin{equation}
h_{\mu c} = e_{\mu}^a h_{ac} = (\delta_{\mu}^a - \frac{1}{2} h_{\mu}^a)
\end{equation}

For further corrections to the first order, we can write more precisely the components of the metric and curvature. Now, we define the components:

\begin{equation}
h_{0c}(x) = \frac{1}{3} \int dx_b \left( \int d^4 x' G(x,x') \, \varepsilon_{bck} (\bar{\psi}_L \sigma^k \psi_L - \bar{\psi}_R \sigma^k \psi_R) \right)
\end{equation}

\begin{equation}
h_{ac}(x) = -\frac{1}{3} \int dx_0 \left( \int d^4 x' G(x,x') \, \varepsilon_{ack} (\bar{\psi}_L \sigma^k \psi_L - \bar{\psi}_R \sigma^k \psi_R) \right)
\end{equation}

\begin{equation}
h_{ac}(x) = -\frac{1}{3} \int d x_b \left( \int d^4 x' G(x,x') \, \varepsilon_{bca} (\bar{\psi}_L \psi_R - \bar{\psi}_R \psi_L) \right)
\end{equation}

The curvature contributions are given by:

\begin{equation}
R_{0c} \approx -\frac{1}{6} \int dx_b \, \varepsilon_{bck} (\bar{\psi}_L \sigma^k \psi_L - \bar{\psi}_R \sigma^k \psi_R)
\end{equation}

\begin{equation}
R_{ac} \approx +\frac{1}{6} \int dx_0 \, \varepsilon_{ack} (\bar{\psi}_L \sigma^k \psi_L - \bar{\psi}_R \sigma^k \psi_R)
\end{equation}

\begin{equation}
R_{ac} \approx +\frac{1}{6} \int d x_b \, \varepsilon_{bca} (\bar{\psi}_L \psi_R - \bar{\psi}_R \psi_L)
\end{equation}

The term \( (\bar{\psi}_L \sigma^k \psi_L - \bar{\psi}_R \sigma^k \psi_R) \) that appears in the source suggests a difference between the two chiral currents. This term is skewed with respect to the left and right components, and therefore the gravitational responses (or curvature) may be different for each current. If the currents of left- and right-handed fermions are treated separately, their influence on gravitational curvature will not be the same, and this could lead to distinct gravitational effects, including the possibility of antigravity in the presence of a proper source configuration.

We associate left-handed fermions with a positive curvature and right-handed fermions with negative curvature. If \( R < 0 \), it implies gravitational repulsion, where geodesics tend to move away on a hyperbolic surface.

Breaking the chiral symmetry could generate a non-symmetric gravitational field, leading to differentiation between the two "directions" (left/right), causing opposite gravitational (antigravitational) effects depending on the specific conditions of the matter involved. Chiral violation leads to asymmetrical curvature, which could manifest itself in anomalous gravitational effects, including antigravity phenomena.

In this model, the attractive nature is associated with left-handed fermions, while anti-gravity is associated with right-handed fermions. In the world we live in, the left component and the attractive gravitational force prevail. If we attribute positive curvature to the left-handed source, the presence of right-handed sources can lead to negative curvature.

This interpretation must be confirmed through gravitational effects. What we can state is that chirality has opposite effects on curvature.
\subsection{Flip between chiralities and Majorana neutrinos}

Consider the following equations involving neutrino contributions to the metric perturbations:

\begin{equation}
h_{0c}(x) = \frac{1}{3} \int dx_b \left( \int d^4x' \, G(x,x') \, \varepsilon_{bck} \left[ \bar{\nu}_L \sigma^k \nu_L - \bar{\nu}_R \sigma^k \nu_R \right] \right)
\end{equation}

\begin{equation}
h_{ac}(x) = -\frac{1}{3} \int dx^0 \left( \int d^4x' \, G(x,x') \, \varepsilon_{ack} \left[ \bar{\nu}_L \sigma^k \nu_L - \bar{\nu}_R \sigma^k \nu_R \right] \right)
\end{equation}

\begin{equation}
h_{ac}(x) = -\frac{1}{3} \int dx_b \left( \int d^4x' \, G(x,x') \, \varepsilon_{bca} \left[ \bar{\nu}_L \nu_R - \bar{\nu}_R \nu_L \right] \right)
\end{equation}

Right-handed neutrinos do not interact via the weak interaction, which selects only left-handed fermions or right-handed antifermions. If right-handed neutrinos exist, they could interact gravitationally and contribute to the perturbation of spacetime curvature, playing a fundamental role in cosmic expansion.

In the Standard Model, neutrinos are strictly left-handed. However, right-handed neutrinos could be obtained via chirality flip of left-handed ones. This flip can occur through a Yukawa-type interaction mediated by a field \( h_{\mu c|b} \), as in the approximate Lagrangian:

\begin{equation}
\mathcal{L}_{h_{\mu c|b}} \approx \frac{1}{4} \partial_\mu h_{\mu c|b} \, \partial^\mu h_{\mu c|b} \, \left\{ \hat{\gamma}^\mu, \hat{\sigma}^{bc} \right\}^2 - \mu^2 \left( h_{\mu c|b} \left\{ \hat{\gamma}^\mu, \hat{\sigma}^{bc} \right\} \right)^2 + \frac{1}{8} h_{\mu c|b} \, \bar{\nu} \left\{ \hat{\gamma}^\mu, \hat{\sigma}^{bc} \right\} \nu
\end{equation}

Now consider neutrinos in the Weyl representation. The chirality flip \( \nu_L \leftrightarrow \nu_R \) leads to the following expressions:

\begin{equation}
\bar{\nu} M^{0bc} \nu = 2 \varepsilon_{bck} \left( \bar{\nu}_L \sigma^k \nu_L - \bar{\nu}_R \sigma^k \nu_R \right)
\end{equation}

\begin{equation}
\nu = \begin{pmatrix} \nu_L \\ \nu_R \end{pmatrix}
\end{equation}

\begin{equation}
2 \bar{\psi} M^{kbc} \psi = 4 \varepsilon_{bck} \left( \bar{\psi}_L \psi_R - \bar{\psi}_R \psi_L \right)
\end{equation}

\begin{equation}
\bar{\nu} M^{kbc} \nu = 2 \varepsilon_{bck} \left( \bar{\nu}_L \nu_R - \bar{\nu}_R \nu_L \right)
\end{equation}

In the weak gravity limit, the perturbation \( h_{ac} \) is practically the same as \( h_{\mu c} \), because tetrads approach the identity matrix. Thus:

\begin{equation}
h_{\mu c|b} \approx h_{a c|b}
\end{equation}

The Yukawa-like couplings that remain in the Lagrangian include:

\begin{equation}
\frac{1}{8} h_{0c|b} \left[ 2 \varepsilon_{bck} \left( \bar{\nu}_L \sigma^k \nu_L - \bar{\nu}_R \sigma^k \nu_R \right) \right]
\end{equation}

This does not directly mix left and right chiralities, but introduces a spinorial coupling:

\begin{equation}
\frac{1}{8} h_{kc|b} \left[ 2 \varepsilon_{bck} \left( \bar{\nu}_L \nu_R - \bar{\nu}_R \nu_L \right) \right]
\end{equation}

This spinorial term describes an interaction between left- and right-handed neutrinos mediated by the gravitational perturbation. This structure implies a chirality flip, where left-handed neutrinos can convert into right-handed ones and vice versa. Such a process resembles a Yukawa-type coupling, where chiral mixing breaks parity symmetry and can lead to mass generation for the neutrino.

If the gravitational field \( h_{\mu c} \) acquires an effective mass or acts as a gravitational analogue of the Higgs field, it could induce spontaneous symmetry breaking in the chiral sector. This mechanism would allow neutrinos to become massive even if they are initially massless, purely through their interaction with gravity.

In this mechanism, neutrinos could undergo chirality flips like Majorana neutrinos, whose left and right components are inherently mixed. For Majorana neutrinos, the spinor can be written as:
\begin{equation}
\nu_M = \begin{pmatrix} \nu_L \\ \nu_R \end{pmatrix}, \quad \nu_R = C \overline{\nu_L}^T,
\end{equation}
where \( C \) is the charge conjugation operator. This relation implies mixing of chiralities unique to Majorana neutrinos.

Majorana neutrinos allow the chirality flip because their mass term mixes chiralities.

This is in contrast with Dirac neutrinos, which, without mass, are purely left-handed and do not allow such flips.
However, if Dirac neutrinos acquire mass, either through conventional mechanisms (seesaw) or via gravitational coupling as in this model, chirality mixing becomes possible, enabling chirality flips similar to those of Majorana neutrinos.
\subsubsection*{Connection to the Majorana Mass Term}

The term
\begin{equation}
\frac{1}{8} h_{k\;c|b} \left[ 2\varepsilon_{bck} \left( \bar{\nu}_L \nu_R - \bar{\nu}_R \nu_L \right) \right]
\end{equation}
in the Lagrangian describes a coupling between chiral neutrino currents and the perturbative gravitational field $h_{\mu\nu}$. This structure is formally analogous to the Majorana mass term
\begin{equation}
\mathcal{L}_M = -\frac{1}{2} m_M \left( \bar{\psi}_L \psi_L^c + \bar{\psi}_R \psi_R^c \right) + \text{h.c.}
\end{equation}
as both express a chirality flip between left- and right-handed components of the neutrino. However, while in the Majorana case lepton number violation is introduced explicitly through a fundamental mass term, in our model the flip arises dynamically as an effect of the interaction with the gravitational field. This suggests a gravitational mechanism for mass generation in neutrinos, where chirality symmetry breaking is induced by the spacetime geometry itself.

\subsection{Flip as scattering}

If we admit the flip of left-handed neutrinos into right-handed neutrinos, we obtain negative curvature and could explain the effects of the expanding universe related to dark matter as antigravitational effects. 
The flip here is due to the interaction between neutrinos and gravity, and can be interpreted as a scattering process described by the integral:
\begin{equation}
h_{ac}(x) = -\frac{1}{3} \int dx_b \left( \int d^4x' \, G(x,x') \, \varepsilon_{bca} \left( \bar{\nu}_L \nu_R - \bar{\nu}_R \nu_L \right) \right)
\end{equation}

Let's consider the two individual terms:
\begin{equation}
h_{ac}(x) = -\frac{1}{3} \int dx_b \left( \int d^4x' \, G(x,x') \, \varepsilon_{bck} \, \bar{\nu}_L \nu_R \right)
\end{equation}
\begin{equation}
h_{ac}(x) = +\frac{1}{3} \int dx_b \left( \int d^4x' \, G(x,x') \, \varepsilon_{bck} \, \bar{\nu}_R \nu_L \right)
\end{equation}

The first process represents a flip of chirality mediated by the field \( h_{\mu c} \). A left-handed neutrino enters, couples with the gravitational field, and produces a right-handed neutrino:
\[
\nu_L \rightarrow h_{\mu\nu} + \nu_R
\]
\begin{figure}[H]
    \centering
    \includegraphics[width=0.3\linewidth]{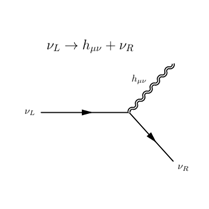}
    \caption{Neutrino L-R.}
    \label{fig:neutrino-lr}
\end{figure}

The second term describes an incoming right-handed neutrino that emits a field and turns into a left-handed neutrino:
\[
\nu_R \rightarrow h_{\mu\nu} + \nu_L
\]
\begin{figure}[H]
    \centering
    \includegraphics[width=0.3\linewidth]{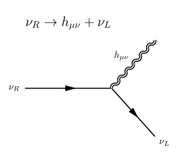}
    \caption{Neutrino R-L.}
    \label{fig:neutrino-rl}
\end{figure}

The Standard Model only predicts the presence of left-handed neutrinos, so initially we consider only the flip from left-handed to right-handed.

The equation
\begin{equation}
h_{ac}(x) = -\frac{1}{3} \int dx_b \left( \int d^4x' \, G(x,x') \, \varepsilon_{bca} \left( \bar{\nu}_L \nu_R - \bar{\nu}_R \nu_L \right) \right)
\end{equation}
leads to the creation of right-handed neutrinos via the flip. If we consider Majorana neutrinos, right-handed neutrinos are left-handed antineutrinos with opposite chirality, so lepton number is not conserved.

\subsection{Calculation of Neutrino-Induced Curvature}

Once right-handed neutrinos are introduced, they contribute to the curvature of spacetime and can potentially induce negative curvature. Starting from the metric perturbations:

\begin{equation}
h_{0c}(x) = \frac{1}{3} \int dx_b \left( \int d^4x' \, G(x,x') \, \varepsilon_{bck} \left[ \bar{\nu}_L \sigma^k \nu_L - \bar{\nu}_R \sigma^k \nu_R \right] \right),
\end{equation}

\begin{equation}
h_{ac}(x) = -\frac{1}{3} \int dx^0 \left( \int d^4x' \, G(x,x') \, \varepsilon_{ack} \left[ \bar{\nu}_L \sigma^k \nu_L - \bar{\nu}_R \sigma^k \nu_R \right] \right),
\end{equation}

we find that the inclusion of right-handed neutrinos reduces the total curvature. The gravitational field distinguishes between left- and right-handed currents, which contribute with opposite signs due to the antisymmetric structure of the Levi-Civita tensor. If we allow for chirality-flipping transitions, a right-handed neutrino contribution corresponds to an antigravitational effect.

The components of the Ricci tensor can be approximated by:

\begin{equation}
R_{0c} \simeq -\frac{1}{6} \int dx_b \, \varepsilon_{bck} \left( \bar{\nu}_L \sigma^k \nu_L - \bar{\nu}_R \sigma^k \nu_R \right),
\end{equation}

\begin{equation}
R_{0c}^{(L)} \simeq -\frac{1}{6} \int dx_b \, \varepsilon_{bck} \left( \bar{\nu}_L \sigma^k \nu_L \right),
\end{equation}

\begin{equation}
R_{0c}^{(R)} \simeq +\frac{1}{6} \int dx_b \, \varepsilon_{bck} \left( \bar{\nu}_R \sigma^k \nu_R \right),
\end{equation}

\begin{equation}
R_{ac} \simeq +\frac{1}{6} \int dx_0 \, \varepsilon_{ack} \left( \bar{\nu}_L \sigma^k \nu_L - \bar{\nu}_R \sigma^k \nu_R \right),
\end{equation}

\begin{equation}
R_{ac}^{(L)} \simeq \frac{1}{6} \int dx_0 \, \varepsilon_{ack} \left( \bar{\nu}_L \sigma^k \nu_L \right),
\end{equation}

\begin{equation}
R_{ac}^{(R)} \simeq -\frac{1}{6} \int dx_0 \, \varepsilon_{ack} \left( \bar{\nu}_R \sigma^k \nu_R \right).
\end{equation}

These contributions from left- and right-handed neutrinos have direct implications for cosmology. The left-handed components increase the curvature, producing an attractive gravitational effect analogous to what is attributed to dark matter. This will be further explored in relation to gravitational lensing and galactic rotation curves.

In contrast, right-handed neutrinos (or equivalently, left-handed antineutrinos) reduce the curvature. This negative contribution may be related to the accelerated expansion of the universe and therefore associated with dark energy.

A nontrivial contribution to curvature appears when both chiralities are present, represented by the mixed term:

\begin{equation}
R_{ac} \simeq \frac{1}{6} \int dx_b \, \varepsilon_{bca} \left( \bar{\nu}_L \nu_R - \bar{\nu}_R \nu_L \right),
\end{equation}

which emerges directly from chirality-flip processes (i.e., $\nu_L \rightarrow \nu_R$), made possible by the neutrino mass. These terms vanish in the absence of right-handed neutrinos.

In such a case, curvature is governed only by the purely left-handed components:

\begin{equation}
R_{0c}^{(L)} \simeq -\frac{1}{6} \int dx_b \, \varepsilon_{bck} \left( \bar{\nu}_L \sigma^k \nu_L \right),
\end{equation}

\begin{equation}
R_{ac}^{(L)} \simeq \frac{1}{6} \int dx_b \, \varepsilon_{ack} \left( \bar{\nu}_L \sigma^k \nu_L \right),
\end{equation}

which induce a positive curvature of spacetime, leading to gravitational attraction similar to the conventional role of dark matter.

However, even a partial inclusion of right-handed neutrinos introduces a significant mixed contribution, represented for instance by:

\begin{equation}
R_{ac}^{(LR)} \simeq \frac{1}{6} \int dx_b \, \varepsilon_{bca} \left( \bar{\nu}_L \nu_R \right),
\end{equation}

This term expresses the curvature associated with the transition $\nu_L \rightarrow \nu_R$ and may involve complex contributions related to chirality interference. Such processes could be relevant for spontaneous Lorentz symmetry breaking and act as sources of torsion in spacetime.

These effects, albeit difficult to interpret in terms of real curvature due to their complex structure, may be essential in describing early-universe dynamics or non-standard gravitational phases. Their natural appearance is more compatible with a Majorana nature of neutrinos, although Dirac neutrinos with mass and chiral coupling may also accommodate them.

This model offers a novel interpretation of how left- and right-handed neutrinos influence the evolution of the universe, with possible implications for understanding both dark matter and dark energy. Future analyses will examine lensing effects and galaxy rotation curves to test the compatibility of this framework with observations.
\subsection{Oriented Spin Currents - Coherence Domains}

In order to better understand the physical interpretation of the induced curvature, we proceed with a detailed examination of the governing equations. The Levi-Civita tensor introduces sign differences between left- and right-handed contributions, thereby inducing opposite modifications in the curvature tensor. We define the spin currents \( J_L^k = (\bar{v}_L) \sigma^k v_L \) and \( J_R^k = (\bar{v}_R) \sigma^k v_R \).

The curvature components are given by:

\begin{equation}
R_{0c} \simeq -\frac{1}{6} \int dx_b \, \varepsilon_{bck} \left( J_L^k - J_R^k \right),
\end{equation}

\begin{equation}
R_{ac} \simeq +\frac{1}{6} \int dx_0 \, \varepsilon_{ack} \left( J_L^k - J_R^k \right),
\end{equation}

\noindent where the left-handed neutrino contributions are:

\begin{equation}
R_{0c}^{(L)} \simeq -\frac{1}{6} \int dx_b \, \varepsilon_{bck} J_L^k,
\end{equation}

\begin{equation}
R_{ac}^{(L)} \simeq +\frac{1}{6} \int dx_0 \, \varepsilon_{ack} J_L^k,
\end{equation}

\noindent and the right-handed neutrino contributions are:

\begin{equation}
R_{0c}^{(R)} \simeq +\frac{1}{6} \int dx_b \, \varepsilon_{bck} J_R^k,
\end{equation}

\begin{equation}
R_{ac}^{(R)} \simeq -\frac{1}{6} \int dx_0 \, \varepsilon_{ack} J_R^k.
\end{equation}

In the absence of coherent alignment in the left-handed neutrino currents, we expect that the contributions to both the temporal and spatial curvature may become disordered or chaotic. The lack of an organized spinor current structure implies that gravitational effects from left-handed neutrinos will not manifest as a uniform attraction or repulsion on cosmological scales, and the curvature contributions will vary locally, influenced by the distribution and interactions of neutrinos. In analogy with Cooper pairs in superconductivity, we hypothesize that left-handed neutrinos can form coherent pairwise structures mediated by torsion and chirally induced gravitational interactions. This coherence leads to a statistically preferred orientation of the spinor current \( J_L^k \), generating a net positive temporal curvature \( R_{0c} > 0 \) and positive spatial curvature \( R_{ac} > 0 \). Such curvature enhances gravitational attraction and supports the interpretation of left-handed neutrinos as a candidate for dark matter. 

The coherent alignment of the right-handed spinor current \( J_R^k \) induces a negative temporal curvature \( R_{0c} < 0 \) and a negative spatial curvature \( R_{ac} < 0 \), resulting in a net repulsive gravitational effect. This geometric contribution is consistent with the behavior expected from dark energy, suggesting that right-handed neutrinos could play a fundamental role in driving the accelerated expansion of the universe. The macroscopic coherence of these spinor currents introduces anisotropies in spacetime that manifest only on large, cosmological scales.

The creation of coherence domains with precise alignments of currents could be due to the interaction of neutrinos with the massive graviton emitted during the flip. Due to torsion, neutrinos interact gravitationally with each other via massive gravitons, creating a network of neutrino-coherent currents. The coherence of these currents, and the resulting gravitational effect, give rise to a curvature that, on a large scale, could behave as a form of dark matter or dark energy, depending on the nature of the neutrinos.The chiral coupling between spinors and gravitation, by torsion and massive graviton, can generate coherent curvatures on a cosmological scale, with direct applications on dark matter (left-handed neutrinos) and dark energy (right-handed).
\subsection{Yukawa Interaction and Gravitational Meissner Effects}

Since we get massive gravity, we can realize a Higgs-like mechanism that modifies fermionic mass. In the Standard Model, the Higgs boson gives mass to fermions through the Yukawa interaction; here we assume a similar situation: gravity interacts with the fermionic current as shown by the interaction term

\begin{equation}
\mathcal{L}_I = -\frac{1}{8} h_{\mu\;c|b} \left[ \bar{\psi} \left\{ \hat{\gamma}^\mu, \hat{\sigma}^{bc} \right\} \psi \right]
\end{equation}

We compare this with the standard Yukawa interaction:

\begin{equation}
\mathcal{L}_y = -k \bar{\psi} \phi \psi
\end{equation}

We approximate the mass starting from the expression:

\begin{equation}
m_\psi \approx k \left\langle h_{\mu\;c|b} \right\rangle
\end{equation}

and assume the background value of the field is determined by the symmetry breaking:

\begin{equation}
\left\langle h_{\mu\;c|b} \right\rangle = \pm \sqrt{ \frac{-\mu^2}{2\lambda} }
\end{equation}

Thus, the total fermion mass becomes:

\begin{equation}
m_{\text{tot}} = m_{\text{fermion}} + k m_\psi
\end{equation}

Fermions interact with the field \( h_{\mu\;c|b} \) through torsion and acquire additional mass. The total mass results from the sum of the intrinsic mass and the mass acquired during the interaction. Here, \( k \) is a coupling constant of the Yukawa-type interaction, and \( \left\langle h_{\mu\;c|b} \right\rangle \) is the background value of the field that depends on the symmetry breaking.

We consider the Yukawa potential:

\begin{equation}
V(r) \approx -\frac{e^{-mr}}{r}
\end{equation}

When the field \( h_{\mu\;c|b} \) interacts with spin particles, the interaction modifies the strength of the gravitational field. Over very large distances, this gravitational field becomes negligible much earlier than Newton's law and doesn’t modify Newtonian gravity. 

Near the source, the exponential \( e^{-mr} \) modulates the rate at which the potential grows, but in a less drastic way than Newton’s law. The interaction with torsion creates a screen to gravity—a confined region where gravity doesn’t propagate—similar to a Meissner effect, like how electromagnetic fields are expelled in a superconductor.

This effect is due to the fermion spin current, creating a “repulsive” gravitational field.

\section{Summary and Conclusion}

The proposed model is based on the interaction between spinors and gravity in the presence of torsion, which arises naturally from fermionic spin currents. The starting point is the Lagrangian for fermions in weak gravity, where the spin connection introduces interaction terms such as:

\begin{equation}
\mathcal{L}_I = -\frac{1}{8} h_{\mu\;c|b} \left[ \bar{\psi} \left\{ \hat{\gamma}^\mu, \hat{\sigma}^{bc} \right\} \psi \right]
+ \frac{1}{16} h_a^\mu h_{\mu\;c|b} \left[ \bar{\psi} \left\{ \hat{\gamma}^a, \hat{\sigma}^{bc} \right\} \psi \right]
\end{equation}

A Ginzburg-Landau-type potential is introduced for the perturbative gravitational field:

\begin{equation}
V\left(h_{\mu\;c|b} \left\{ \hat{\gamma}^\mu, \hat{\sigma}^{bc} \right\}\right) 
= \mu^2 \left( h_{\mu\;c|b} \left\{ \hat{\gamma}^\mu, \hat{\sigma}^{bc} \right\} \right)^2 
+ \lambda \left( h_{\mu\;c|b} \left\{ \hat{\gamma}^\mu, \hat{\sigma}^{bc} \right\} \right)^4
\end{equation}

When \(\mu^2 < 0\) and \(\lambda > 0\) this potential leads to a non-zero vacuum expectation value and implies spontaneous Lorentz symmetry breaking.

\begin{equation}
\left\langle h_{\mu\;c|b} \left\{ \hat{\gamma}^\mu, \hat{\sigma}^{bc} \right\} \right\rangle = \pm \sqrt{\frac{-\mu^2}{2\lambda}}
\end{equation}

This non-zero minimum is directly related to the fermionic current that emerges naturally in general relativity in the presence of spinors.

The most direct and significant consequence of the breaking of the Lorentz symmetry, assimilated to the Ginzburg-Landau/Higgs mechanism, is that gravity acquires mass.

We calculate the field equation for the perturbation field \(h_{\mu\;c|b}\) which then satisfies the Klein-Gordon equation:

\begin{equation}
\square \left(h_{\mu\;c|b}\right) + 8 \mu^2 h_{\mu\;c|b} = 0
\end{equation}

Indicating massive gravity with mass \(\sqrt{8}\, m\).

In addition, fermions can acquire mass through their interaction with the gravitational field. These mass states may be highly unstable and decay rapidly, making them effectively undetectable like dark matter. 

The massive nature of gravity implies short-range behavior, the mass acquired by the gravitational field implies that the interaction mediated by this field does not propagate indefinitely, but decays exponentially with distance. This makes the interaction short-range and confines it to specific regions. This phenomenon is described as a kind of "gravitational shielding" or gravitational Meissner effect, where gravity is "ejected" or attenuated in regions where the spinors couple strongly with gravity due to the fermionic spin current.

In summary, Lorentz symmetry breaking, driven by torsion-mediated spin-gravity interaction, transforms the nature of gravity from a massless, long-range interaction to a massive, short-range interaction, offering a new framework for interpreting cosmological phenomena. 

Starting from the metric perturbation, the curvature is analyzed. Torsion-induced couplings between spinor currents and the gravitational field generate curvature through chirality-sensitive terms which contribute to the metric perturbation. In this picture, left-handed fermions generate attractive gravity, while right-handed fermions induce repulsive curvature. This mechanism offers a geometric explanation for galactic rotation curves and the cosmic acceleration usually attributed to dark matter and dark energy.

A key role is played by Majorana neutrinos, whose chirality-flipping processes are mediated by the gravitational field, interpreted here as a dynamical scattering mechanism that violates lepton number. 

Right-handed neutrinos, which are sterile under Standard Model interactions, may contribute gravitationally as a source of repulsive curvature, mimicking dark energy.

In order to produce observable gravitational effects over cosmological distances, the neutrino currents cannot remain randomly oriented at microscopic scales. Instead, they must organize into coherence domains—spatial regions where the chiral neutrino currents are preferentially aligned along certain directions. This alignment is crucial because, without it, the local contributions of individual neutrinos to curvature or torsion would average out and vanish on large scales. Within each coherence domain, the constructive interference of neutrino currents amplifies their cumulative gravitational effect, enabling the emergence of macroscopic phenomena such as spacetime curvature. The formation of these domains may be triggered by mechanisms such as chiral flips, interactions with a background field, or phase transitions in the early universe.

Cosmological observations indicate the presence of dark matter, responsible for the anomalous rotational dynamics of galaxies, and dark energy, which drives the accelerated expansion of the universe. However, the microscopic origin of both remains not fully explained. In this framework, gravity is mediated by massive gravitons, and its interaction with fermionic spin—particularly through torsion—modifies the nature of gravity itself, making it effectively short-ranged and chiral. This offers a possible alternative to traditional dark matter and the cosmological constant, associating the observed effects with the dynamics of spinor currents and symmetry breaking in curved spacetime.

In essence, the theory proposes that the intrinsic spin of fermions, through its coupling with torsion and gravity, modifies the structure of spacetime itself. This leads to mass acquisition for both fermions and gravitons, gravitational shielding, and chirality-dependent curvature effects, providing a unified and testable explanation for gravitational lensing, galaxy rotation curves, and cosmic acceleration.

\appendix
\section{Coherence domains}
We study in detail the pattern of the neutrino currents required to generate attractive and repulsive gravity.

\subsection{Left neutrinos contributions:}
\begin{align}
R_{0cL} &\approx -\frac{1}{6} \int dx_b \, \epsilon_{bck} J_L^k > 0 \\
R_{01L} &\approx -\frac{1}{6} \int dx_b \, \epsilon_{b1k} J_L^k = -\frac{1}{6} \int dx_2 \, \epsilon_{213} J_L^3 - \frac{1}{6} \int dx_3 \, \epsilon_{312} J_L^2 \\
&= \frac{1}{6} \int dx_2 J_L^3 - \frac{1}{6} \int dx_3 J_L^2 \\
R_{02L} &\approx -\frac{1}{6} \int dx_b \, \epsilon_{b2k} J_L^k = -\frac{1}{6} \int dx_1 \, \epsilon_{123} J_L^3 - \frac{1}{6} \int dx_3 \, \epsilon_{321} J_L^1 \\
&= -\frac{1}{6} \int dx_1 J_L^3 + \frac{1}{6} \int dx_3 J_L^1 \\
R_{03L} &\approx -\frac{1}{6} \int dx_b \, \epsilon_{b3k} J_L^k = -\frac{1}{6} \int dx_1 \, \epsilon_{132} J_L^2 - \frac{1}{6} \int dx_2 \, \epsilon_{231} J_L^1 \\
&= \frac{1}{6} \int dx_1 J_L^2 - \frac{1}{6} \int dx_2 J_L^1
\end{align}

\begin{align}
\int dx_2 J_L^3 &> \int dx_3 J_L^2 \\
\int dx_1 J_L^3 &< \int dx_3 J_L^1 \\
\int dx_1 J_L^2 &> \int dx_2 J_L^1
\end{align}

\begin{align}
R_{acL} &\approx \frac{1}{6} \int dx_0 \, \epsilon_{ack} J_L^k > 0 \\
R_{12L} &\approx \frac{1}{6} \int dx_0 \, \epsilon_{123} J_L^3 = \frac{1}{6} \int dx_0 J_L^3 \\
R_{13L} &\approx \frac{1}{6} \int dx_0 \, \epsilon_{132} J_L^2 = -\frac{1}{6} \int dx_0 J_L^2 \\
R_{32L} &\approx \frac{1}{6} \int dx_0 \, \epsilon_{321} J_L^1 = -\frac{1}{6} \int dx_0 J_L^1
\end{align}

\begin{align}
\int dx_0 J_L^3 &> 0 \\
\int dx_0 J_L^2 &< 0 \\
\int dx_0 J_L^1 &< 0
\end{align}

\subsection{Right neutrinos contributions:}
\begin{align}
R_{0cR} &\approx \frac{1}{6} \int dx_b \, \epsilon_{bck} J_R^k < 0 \\
R_{01R} &\approx \frac{1}{6} \int dx_b \, \epsilon_{b1k} J_R^k = \frac{1}{6} \int dx_2 \, \epsilon_{213} J_R^3 + \frac{1}{6} \int dx_3 \, \epsilon_{312} J_R^2 \\
&= -\frac{1}{6} \int dx_2 J_R^3 + \frac{1}{6} \int dx_3 J_R^2 \\
R_{02R} &\approx \frac{1}{6} \int dx_b \, \epsilon_{b2k} J_R^k = \frac{1}{6} \int dx_1 \, \epsilon_{123} J_R^3 + \frac{1}{6} \int dx_3 \, \epsilon_{321} J_R^1 \\
&= \frac{1}{6} \int dx_1 J_R^3 - \frac{1}{6} \int dx_3 J_R^1 \\
R_{03R} &\approx \frac{1}{6} \int dx_b \, \epsilon_{b3k} J_R^k = \frac{1}{6} \int dx_1 \, \epsilon_{132} J_R^2 + \frac{1}{6} \int dx_2 \, \epsilon_{231} J_R^1 \\
&= -\frac{1}{6} \int dx_1 J_R^2 + \frac{1}{6} \int dx_2 J_R^1
\end{align}

\begin{align}
\int dx_2 J_R^3 &> \int dx_3 J_R^2 \\
\int dx_1 J_R^3 &< \int dx_3 J_R^1 \\
\int dx_1 J_R^2 &> \int dx_2 J_R^1
\end{align}

\begin{align}
R_{acR} &\approx -\frac{1}{6} \int dx_0 \, \epsilon_{ack} J_R^k < 0 \\
R_{12R} &\approx -\frac{1}{6} \int dx_0 \, \epsilon_{123} J_L^3 = -\frac{1}{6} \int dx_0 J_R^3 \\
R_{13R} &\approx -\frac{1}{6} \int dx_0 \, \epsilon_{132} J_L^2 = \frac{1}{6} \int dx_0 J_R^2 \\
R_{32R} &\approx -\frac{1}{6} \int dx_0 \, \epsilon_{321} J_L^1 = \frac{1}{6} \int dx_0 J_R^1
\end{align}

\begin{align}
\int dx_0 J_R^3 &> 0 \\
\int dx_0 J_R^2 &< 0 \\
\int dx_0 J_R^1 &< 0
\end{align}

These constraints indicate that, in the presence of torsion and chiral flip, left-handed neutrinos tend to align along directions that generate \( R_{0cL} > 0 \) and \( R_{acL} > 0 \), which can be associated with the dark matter effect. Right-handed neutrinos, on the other hand, organize themselves according to domains that lead to \( R_{0cR} < 0 \) and \( R_{acR} < 0 \), in accordance with gravitational repulsive behavior, i.e. dark energy.This interpretation is in agreement with the geodesic focusing theorem (Raychaudhuri’s equation) [4].
Coherence arises from the gravitational mediation of the flip: each flip creates a massive graviton that locally distorts space-time, orienting adjacent spinors according to preferred symmetries. The result is the formation of coherence domains, i.e. spatial regions in which neutrinos tend to have similar current orientations.

\section{Fermionic Current}

In the previous study we have seen that in general relativity, torsion arises related to a fermionic current [5]. We consider the following expression:

\begin{equation}
\frac{1}{8} \bar{\psi} e_c^\mu M^{cab} \omega_{\mu ab} \psi
\end{equation}

where \( M^{cab} = [\gamma^c, \sigma^{ab}] \).

In weak gravity, this term becomes:

\begin{equation}
e_c^\mu = \delta_c^\mu - \frac{1}{2} h_c^\mu
\end{equation}

and

\begin{equation}
\omega_{\mu ab} = \frac{1}{2} \left( h_{\mu a|b} - h_{\mu b|a} \right) = \frac{1}{2} \left( \partial_b h_{\mu a} - \partial_a h_{\mu b} \right)
\end{equation}

Substituting these into the previous expression gives:

\begin{equation}
\frac{1}{8} \bar{\psi} e_c^\mu M^{cab} \omega_{\mu ab} \psi \approx \frac{1}{16} \bar{\psi} M^{\mu ab} \left( h_{\mu a|b} - h_{\mu b|a} \right) \psi
\end{equation}

Since the only non-vanishing elements of \( M^{cab} \) are \( M^{0ij}, M^{nij}, M^{i0j} \), we only consider contributions arising from the terms:

\begin{equation}
\frac{1}{8} \bar{\psi} e_0^\mu \omega_{\mu ij} M^{0ij} \psi \approx \frac{1}{16} \bar{\psi} M^{0ij} \left( h_{\mu i|j} - h_{\mu j|i} \right) \psi
\end{equation}

The only non-vanishing terms of \( M^{abc} = [\gamma^a, \sigma^{bc}] \) are:

\begin{equation}
M^{0bc} = 2 \epsilon_{bck} \begin{pmatrix} \sigma^k & 0 \\ 0 & -\sigma^k \end{pmatrix}
\end{equation}

\begin{equation}
M^{b0c} = -M^{0bc} = -2 \epsilon_{bck} \begin{pmatrix} \sigma^k & 0 \\ 0 & -\sigma^k \end{pmatrix}
\end{equation}

\begin{equation}
M^{kbc} = 2 \epsilon_{bck} \begin{pmatrix} 0 & I \\ -I & 0 \end{pmatrix}
\end{equation}

Explicitly speaking about the dynamical arising of torsion, we find this fermionic spin current:

\begin{equation}
\pi_a^\alpha = \frac{\delta S_f}{\delta \partial_\tau e_a^\alpha} = \frac{1}{8} \bar{\psi} \left[ e_0^\tau e_i^\alpha M^{0ia} + \frac{1}{2} e_i^\tau e_0^\alpha M^{i0a} + e_n^\tau e_i^\alpha M^{nia} \right] \psi
\end{equation}

where the index \( a \) is flat spacelike. Also, for the component \( \pi_0^\alpha \), we have:

\begin{equation}
\pi_0^\alpha = \frac{\delta S_f}{\delta \partial_\tau e_0^\alpha} = \frac{1}{8} \bar{\psi} \left[ -\frac{1}{2} e_i^\tau e_j^\alpha M^{i0j} \right] \psi
\end{equation}

\subsection{Fermionic Current in Weak Gravity}

We consider the tetrads expressed in weak gravity and calculate the momentum, eliminating second-order contributions:

\begin{equation}
\pi_a^\alpha \approx \frac{1}{8} \bar{\psi} \left[ \delta_0^\tau \delta_i^\alpha M^{0ia} + h_0^\tau \delta_i^\alpha M^{0ia} + \delta_0^\tau h_i^\alpha M^{0ia} + \frac{1}{2} \delta_i^\tau \delta_0^\alpha M^{i0a} + \delta_n^\tau \delta_i^\alpha M^{nia} \right] \psi
\end{equation}

where the index \( a \) is flat spacelike and the index 0 is flat timelike. The equation of angular momentum reveals the presence of the fermionic spin current, and torsion arises dynamically due to fermions. We see the coupling with gravity described by \( h_0^\tau \) and \( h_i^\alpha \).

For the component \( \pi_0^\alpha \), we have:

\begin{equation}
\pi_0^\alpha \approx -\frac{1}{16} \bar{\psi} \left[ \delta_i^\tau \delta_j^\alpha M^{i0j} + \delta_i^\tau h_j^\alpha M^{i0j} + h_i^\tau \delta_j^\alpha M^{i0j} \right] \psi
\end{equation}

where \( i \) and \( j \) are flat spacelike indices.
\section*{Data Availability}

I have no data associated with the manuscript, as the article is about theoretical analysis.

\section*{Acknowledgments}

I wish to thank Francisco Bulnes for his invaluable support and insights throughout the course of this work.

\section*{References}

\begin{enumerate}
    \item E. Varani, ``Spinor framework in linearized gravity: a perspective from the energy momentum tensor," \textit{Transnational Journal of Mathematical Analysis and Applications}, Vol. 12, Issue 1, 2024, Pages 1-52.
    \item F. Bulnes, ``Deep study of the Universe through torsion," Cambridge Scholars Publishing, UK, 2022.
    \item F. Bulnes, ``Detection and Measurement of Quantum Gravity by a Curvature Energy Sensor: H-States of Curvature Energy," \textit{Recent Studies in Perturbation Theory}, IntechOpen, 2017, Pages 953-513-261-X, 978-953-513-261-5, DOI:10.5772/6802.
    \item A. K. Raychaudhuri, “Relativistic Cosmology. I”
    \textit{Physical Review }(1955)  Vol.98, Pages 1123–1126.
    \item E. Varani, ``Fermionic current in general relativity," \textit{Transnational Journal of Mathematical Analysis and Applications}, Vol. 11, Issue 2, 2023, Pages 89-101.
\end{enumerate}

\end{document}